# Field induced density wave in the heavy fermion compound CeRhIn$_5$

P.J.W. Moll$^{1*}$, B. Zeng$^2$, L. Balicas$^2$, S. Galeski$^1$, F.F. Balakirev$^3$, E.D. Bauer$^4$, F. Ronning$^4$

$^1$Solid State Physics Laboratory, Department of Physics, ETH Zurich, CH-8093 Zurich, Switzerland.
$^2$National High Magnetic Field Laboratory, Florida State University, Tallahassee, FL 32310, USA.
$^3$National High Magnetic Field Laboratory, LANL, E536, Los Alamos, NM 87545, USA.
$^4$Los Alamos National Laboratory, Los Alamos, NM 87545, USA.187

**Metals containing Ce often show strong electron correlations due to the proximity of the 4f state to the Fermi energy, leading to strong coupling with the conduction electrons. This coupling typically induces a variety of competing ground states, including heavy-fermion metals, magnetism and unconventional superconductivity[1]. The d-wave superconductivity in CeTMIn$_5$ (TM=Co, Rh, Ir) has attracted significant interest[2–6] due to its qualitative similarity to the cuprate high-T$_c$ superconductors. Here, we show evidence for a field induced phase-transition to a state akin to a density-wave (DW) in the heavy fermion CeRhIn$_5$, existing in proximity to its unconventional superconductivity[7]. The DW state is signaled by a hysteretic anomaly in the in-plane resistivity accompanied by the appearance of non-linear electrical transport at high magnetic fields (>27T), which are the distinctive characteristics of density-wave states[8]. The unusually large hysteresis enables us to directly investigate the Fermi surface of a supercooled electronic system and to clearly associate a Fermi surface reconstruction with the transition. Key to our observation is the fabrication of single crystal microstructures, which are found to be highly sensitive to "subtle" phase transitions involving only small portions of the Fermi surface. Such subtle order might be a common feature among correlated electron systems, and its clear observation adds a new perspective on the similarly subtle CDW state in the cuprates.**

The heavy-fermion superconductor CeRhIn$_5$ received significant attention due to its particularly feature-rich phase diagram: The local-moment anti-ferromagnetism (AFM) is suppressed by moderate pressure and unconventional superconductivity (SC) with a surprisingly high T$_c^{max}$~2.1K[7] appears. Remarkably, a first order phase transition at p$_c$~17kbar quenches the AFM order in zero magnetic field and stabilizes a non-magnetic SC phase at higher pressures. The main result of our study is the identification of a previously undetected state akin to a density-wave above a critical field H$_c$~27T ("DW", Fig. 1a). The SC and the DW phases share intriguing similarities: Both arise in the vicinity of a quantum critical point bounding the AFM phase, and the transition from the AFM metal into both the DW and the SC states appears to be first order[2]. As charge- and spin-density-waves are mainly observed in low-dimensional electronic systems, the density-wave in CeRhIn$_5$ most likely appears due to an instability of a low-dimensional Fermi surface[8,9] that coexist with three-dimensional Fermi surface sheets[10].

Our observation of density waves in close proximity to unconventional d-wave superconductivity in CeRhIn$_5$ is reminiscent of the recent developments in the cuprate high-T$_c$ superconductors: Their particularly complex phase diagram hosts a variety of ground states of correlated electron matter, some coexisting and others competing with high temperature superconductivity[11,12]. One of these states also is the spin- and charge-order, which was initially thought to be a special case for the La-based cuprates such as $La_{2-x}Ba_xCuO_4$[12,13]. Yet recently, fluctuating charge-order was also observed in $YBa_2Cu_3O_{7-\delta}$[11,14] and the electronic structure observed by quantum oscillation measurements agrees with the broken translational symmetry due to charge-order[15]. Moderate magnetic fields can tune this system into a static charge ordered state, identified by X-ray diffraction and NMR experiments[16,17]. While the lattice distortion and the charge modulation may have a profound effect on the electronic structure of the cuprates, by themselves these density waves are subtle features, which could only very recently be identified experimentally. This naturally leads to the question if similarly subtle charge-order in the vicinity of unconventional superconductivity is a common feature among several other classes of materials. Our results thus suggest that the coexistence, or competition, between charger-order and unconventional superconductivity might be a more common feature among strongly correlated systems than previously thought.

The key to detecting this transition in our study is the Focused Ion Beam based fabrication of single crystal microstructures. This technique may become a powerful tool to detect subtle changes in the electronic matter of a wide range of materials (Fig.1b). Details about device fabrication and an overview of the reproducibility of the results are given in the Supplement and in Refs.[18,19].

Figure 2a shows the magnetoresistance for currents flowing along both the in-plane and the inter-plane direction, or $\rho_a(H)$ and $\rho_c(H)$ respectively, at 380mK. These traces immediately highlight the key observation of this study: The in-plane resistivity shows a step-like transition at 34T to a high-resistive state (red), which more than doubles the resistance value. This jump is followed by a region of negative magnetoresistance up to a minimum in resistivity at 42T. Upon decreasing the magnetic field, the resistivity follows an upper hysteresis branch characterized by an enhancement of the quantum oscillatory amplitude until it drops back onto the lower branch (blue) at 28T. This anomalous reduction in amplitude may be related to the previously reported de Haas – van Alphen amplitude anomalies in $CeRhIn_5$ within this field and temperature range[20]. At the same time, the out-of-plane resistivity $\rho_c(H)$ remains featureless, indicating that mainly those electrons responsible for the in-plane transport participate in the transition. The magnetic torque $\vec{\tau} = \vec{M} \times \vec{H}$ was measured simultaneously with the transport samples using a capacitive cantilever technique (Fig. 2b). It shows pronounced de Haas-van Alphen oscillations, again highlighting the crystal quality, but otherwise remains featureless at the fields at which the jump occurs. This indicates that the Ce-4f anti-ferromagnetic order remains unaffected, thus suggesting a phase-transition involving only the charge degrees of freedom. Consequently, it does not lead to a large change in the density of states at the Fermi level as often observed when the Fermi surface is reconstructed by a charge- or spin-density-wave. The contrasts between the subtle signs for a phase transition in the torque measurements on macroscopic crystals and the sharp features in the charge transport measurements on microstructures is remarkable. This suggests that the differences in internal strain and coupling of the sample to the substrate may be the origin of the strong feature at the phase transition. Electrical transport measurements in microstructured single crystals are thus shown to be a sensitive probe for subtle phase transitions that may be difficult to discern in other measurements, and thus may become an important tool to detect phase transitions in other classes of materials.

We interpret our observation as evidence for a field-induced charge-order transition involving in-plane electrons, and which does not significantly affect either the c-direction resistivity or the magnetic torque. One of the hallmarks of charge-order is a pronouncedly non-linear conductivity associated with an additional conduction channel due to the sliding of a density-wave state in parallel to the remaining, un-gapped quasiparticles[8,21]. For small electric fields, Coulomb interactions pin the charge-density-wave (CDW) to crystal defects, and the remaining free quasiparticles constitute the only accessible conduction channel. The pinning potential can be overcome by the application of an electric field beyond a certain threshold value $E_c$, leading to the sliding motion of the density-wave which contributes to the conductivity. At the transition field, we observe the onset of this type of non-linearity (Figure 3). The large hysteresis allows us to make a strong argument: By preparing the system in the lower (blue) and upper (red) branch, we can compare the differential resistance at the exact same conditions of field and temperature. This also allows us to exclude artifacts such as field dependent contact resistances and self-heating, as these extrinsic sources of error do not depend on the hysteretic state of the sample. The non-linearity was observed in the whole experimentally accessible high-field region up to 45T, indicating the persistence of the density-waves above the hysteretic region.

While the low resistance branch (blue) remains ohmic up to high current densities, a pronouncedly non-linear differential resistance is observed in the upper branch (red). A well-defined critical current $I_c$=190µA separates the regions of pinned and sliding density-wave. Using the sample length $l = 34.6 \mu m$ and the low-current resistance $R = 0.277 \Omega$ in the upper branch, one finds a depinning-threshold electric field $E_c$=$RI_c$/l=15.2mV/cm. Such $E_c$ of the order of 10mV/cm are commonly observed in incommensurate CDW systems such as $NbSe_3$ and $TaS_3$[8,9].

The large hysteresis presents us with the rare opportunity to study a "supercooled" electronic system and to compare the system with and without the charge-order under the same thermodynamic conditions. At the field-induced charge-order transition, the Fermi Surface is reconstructed as indicated by a change in the Shubnikov-de Haas frequency spectrum (Figure 4). Besides a change in amplitude owing to the overall increase of the signal at the transition, two high frequencies around 4680T and 5400T appear in the high field state, indicating the appearance of a large Fermi surface cross-section[22]. By using a smaller field window for the frequency analysis of the metastable region only (Figure 4, middle), we can show that these frequencies are absent in the lower and present in the upper branch in the same field region. Since a reconstruction of the Fermi surface indicates the thermodynamic nature of the observed phase-transition, the hysteretic appearance of this additional frequency firmly establishes it as a first-order transition.

Another important aspect of the observed density-wave is to what extent the order includes charge- and spin-degrees of freedom. Conventional incommensurate charge- and spin-density waves (SDW) lead to similar non-linear transport characteristics. SDW-like anti-ferromagnetic transitions typically yield pronounced changes in the anisotropic susceptibility $\chi$[23], which were not detected by our magnetic torque measurements and thus suggest a charge-order. Yet in the presence of strong magnetic fields, the nesting vector **q** will be field dependent due to the Zeeman-splitting of the Fermi surface and a pure CDW will become unstable in a similar way as a superconductor transitioning into an FFLO state. It was shown that in this situation an unconventional density-wave involving charge- and spin-degrees of freedom may emerge as the new ground state at high magnetic fields[24]. Such an unconventional charge-order has been observed in high magnetic fields in low dimensional materials such as $(Per)_2Pt(mnt)_2$, which shows remarkable similarities to our findings[25]. Such density-wave instabilities are well understood in low-dimensional materials arising from the enhancement of the susceptibility upon reducing the dimensionality[26]. Yet $CeRhIn_5$ shows nearly isotropic transport properties in zero field(Figure 1), and further theoretical treatment will be required to understand this phenomenon in the context of low-dimensional Fermi surface sheets coexisting with strongly 3D sheets[10] in the presence of magnetic fields. In addition, the driving force of this instability will have to be identified and the observation of a density-wave emphasizes the need to consider the often neglected role of electron-phonon coupling in heavy fermion systems. Such coupling, however, is a natural consequence of the sensitivity of the ground state to small changes in bond lengths[7].

Given the differences in the electronic structure between $CeRhIn_5$ and $YBa_2Cu_3O_{7-\delta}$, the signatures of a high field induced charge-order in both systems share surprising similarities: There is no clear anomaly at the charge-order transition in $CeRhIn_5$ observed in the torque measurements, contrary to conventional CDW materials such as $NbSe_3$[27]. It remains an open question as to why the CDW in $YBa_2Cu_3O_{7-\delta}$ can be well observed by NMR or X-Ray diffraction experiments, yet heat capacity, resistivity or magnetic torque studies do not reveal any anomaly at the field-induced transition[28]. This may result from very small changes in entropy at the transition or be related to the intrinsic time constants of the different measurement techniques and thus their different sensitivities to fluctuating order. It suggest that other measurement techniques such as high field NMR should be performed in $CeRhIn_5$, to clarify the so far unknown microscopic structure of the density-wave. The absence of a magnetization anomaly indicates that the field induced charge ordered state in both compounds is a similarly subtle order, involving only a relatively small fraction of carriers. This highlights a different aspect of the similarity between cuprates and Ce-based "115" materials: Apparently, there is an almost undetectably small number of electrons in low-dimensional orbits present in both material classes that can drive a phase-transition in moderately high magnetic fields perhaps involving the lattice degrees of freedom, yet without strong influence on the remaining carriers.

# Figures

**Figure 1: Phase diagram of CeRhIn$_5$**

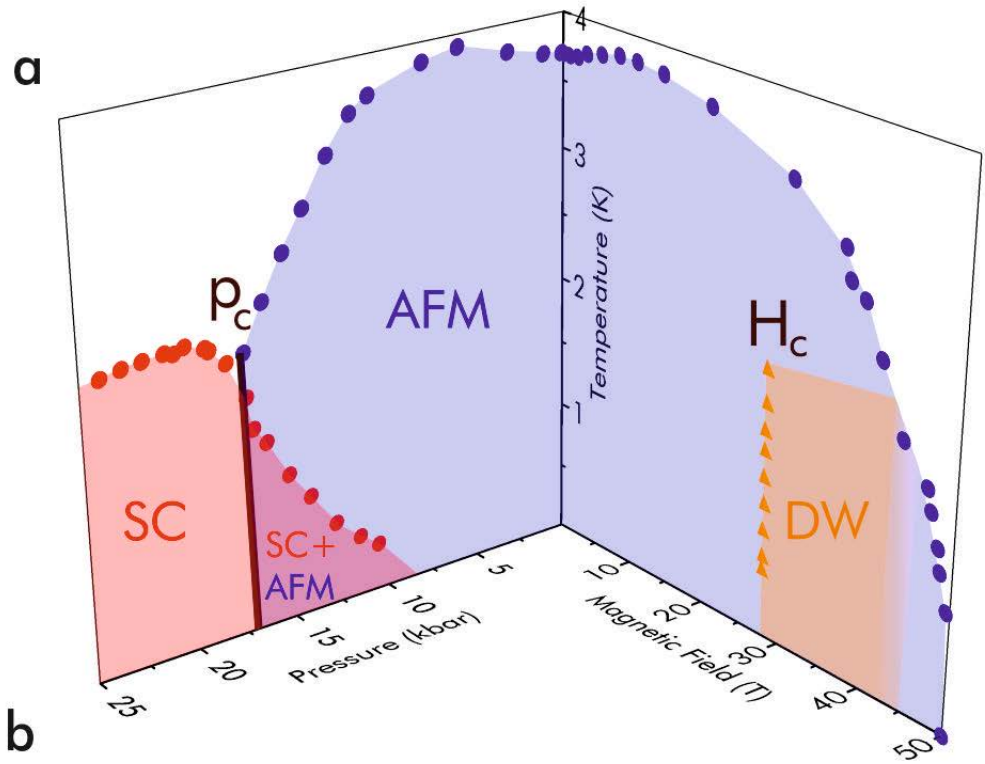

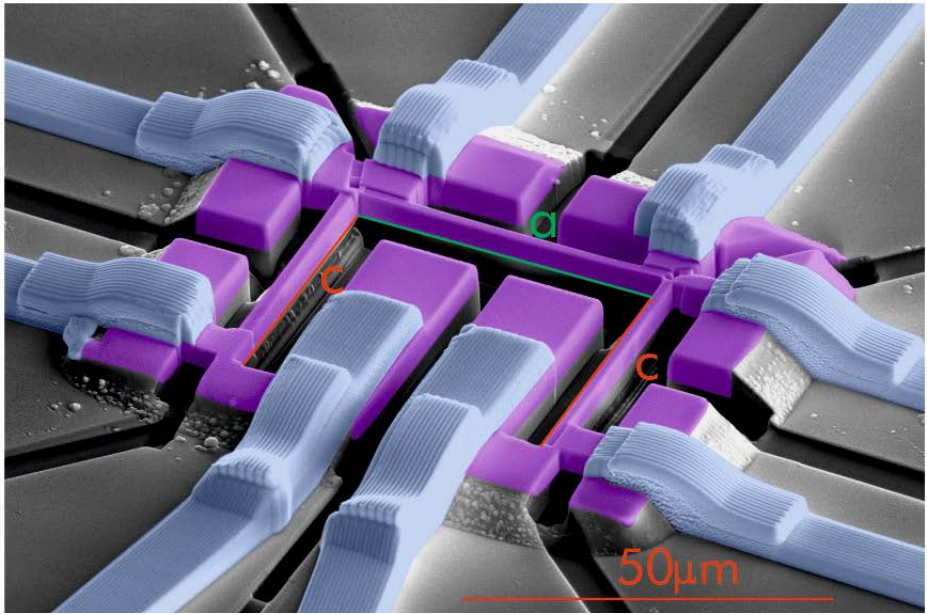

a) Pressure-Field phase diagram of CeRhIn$_5$. At the critical pressure p$_c$, a first order phase transition suppressing the anti-ferromagnetism occurs within the superconducting state. At the critical magnetic field $H_c$, also a first order phase transition into a density-wave state is found, which is the main finding of our study. The transition persists up to temperatures of 2K, above which it becomes unobservable. The DW state clearly coexists with the AFM order, yet it remains an open question if it persists beyond the AFM phase boundary into the paramagnetic state. The (p,T) data is reproduced from ref[29] and $T_N(H)$ from ref[30].

b) SEM micrograph of the FIB carved ~ 60 x 60 µm slice of CeRhIn$_5$ (purple). The (a,c)-plane oriented slice was structured into a U-shaped geometry and contacted by FIB-assisted platinum deposition (blue). Through the lower-left contacts a shared current is applied along the "U", passing through two resistivity bars along the c-direction (red) and one along the a-direction (green). The c-axis bars are both 31.6 µm long with a cross-section of 1.4 µm x 4 µm, and the a-axis bar is 34.5 µm long with a cross-section of 2.2 µm x 4 µm. The high residual resistivity ratios ($\rho_a(300K)/\rho_a(80mK)$~258, $\rho_c(300K)/\rho_c(80mK)$~268), the good quantitative agreement with the temperature dependence of the resistivity on macroscopic crystals[31], and the observation of Shubnikov-de Haas oscillations in fields as low as 4T evidences the high crystal quality of the FIB prepared devices.

**Figure 2: Magnetoresistance and magnetic torque of CeRhIn$_5$ at 350mK**

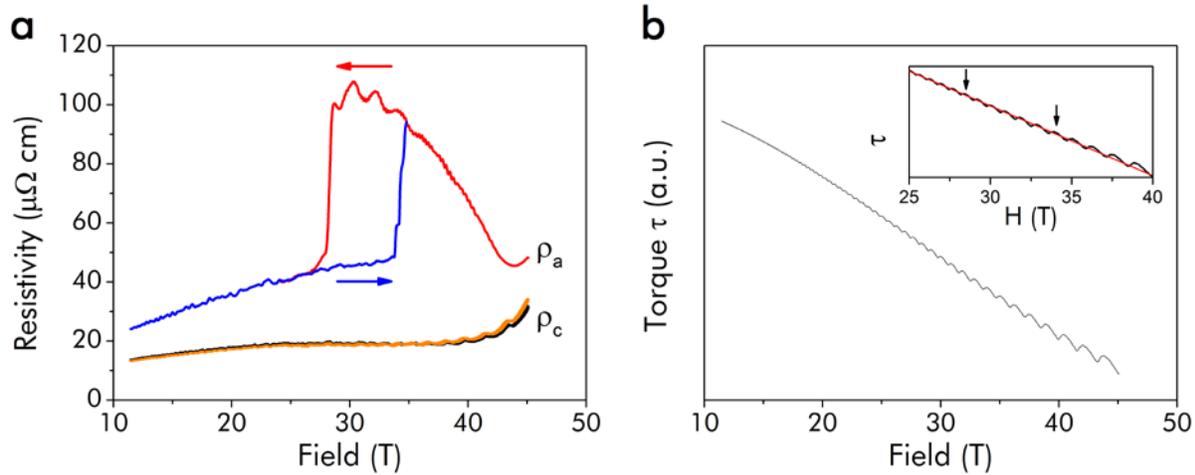

a) Resistivity along the a- and the c-direction as well as b) the magnetic torque of CeRhIn$_5$ single crystals in fields tilted 19° away from the c-axis at 380mK. The two main observations of this study become evident: A hysteretic, first-order like transition in field occurring at ~30T, which is marked by a salient jump in the in-plane resistivity. Remarkably the out-of-plane resistivity as well as the magnetic torque remain featureless across the transition. The inset shows the torque on an amplified scale around the region of the resistive anomaly, with fields of the up- and down-ward resistivity jumps indicated by arrows and a linear fit to the data (red). Except for de Haas – van Alphen oscillations, the torque remains featureless in this region. This suggests a change in two-dimensional pieces of the Fermi Surface prominent in the in-plane transport, but which do not contribute significantly to the out-of-plane transport. At the same time, the featureless torque indicates that a very subtle and practically imperceptible change in the magnetic susceptibility between the two states. This hysteretic feature was observed for a wide range of angles (See Supplement).

**Figure 3: Non-linear transport in the hysteretic region**

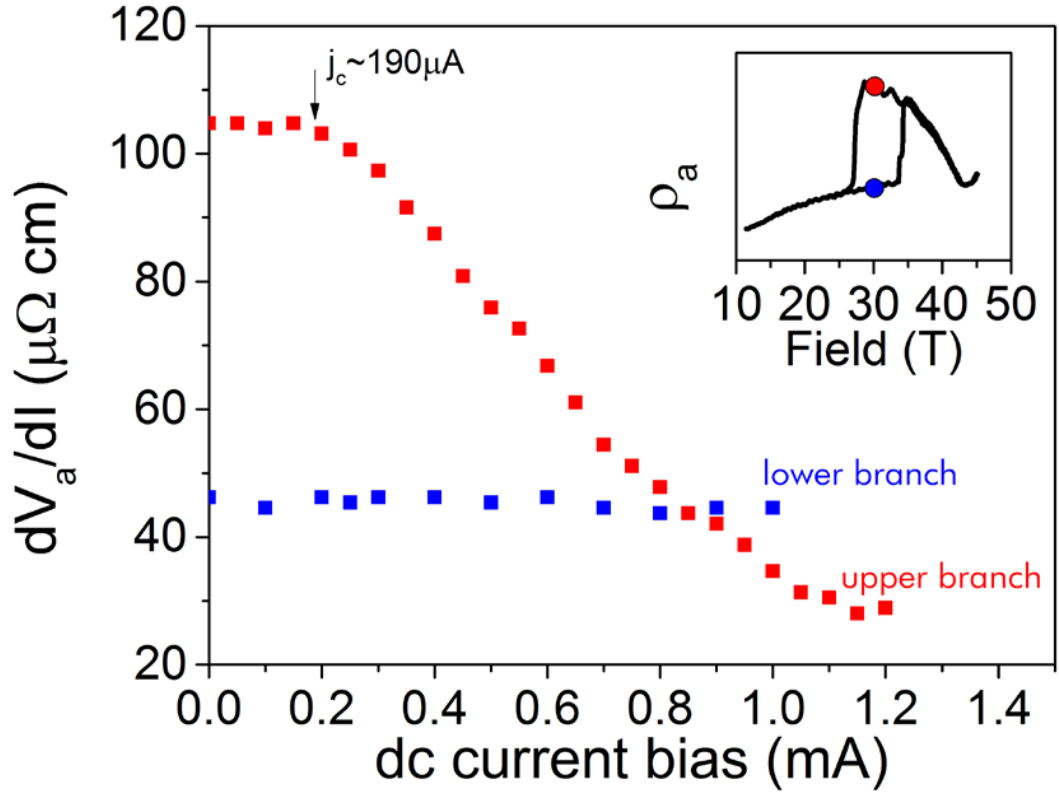

The differential in-plane resistivity $dV_a/dI$, measured by applying a small (10μA RMS) ac-current superimposed onto a dc current bias, highlighting the key observation of this study. Both traces were measured under the same conditions at 30T and 380mK. The system was prepared once in the low-resistance branch (blue), and then cycled to 45T and back to reach the high-resistance branch (red). The zero-bias differential resistance changed by a factor of 2, indicating a successful state preparation. While in the low-resistivity branch the system remained ohmic over a wide range of current bias, the high-resistance branch shows pronouncedly non-linear transport. This behavior is typical for pinned charge density waves, which are depinned by Coulomb forces exceeding the pinning force. In addition, self-heating can be excluded as the additional sample resistance due to the transition is negligible compared to the contact resistance of 40Ω.

**Figure 4: Quantum oscillations indicate a Fermi Surface reconstruction**

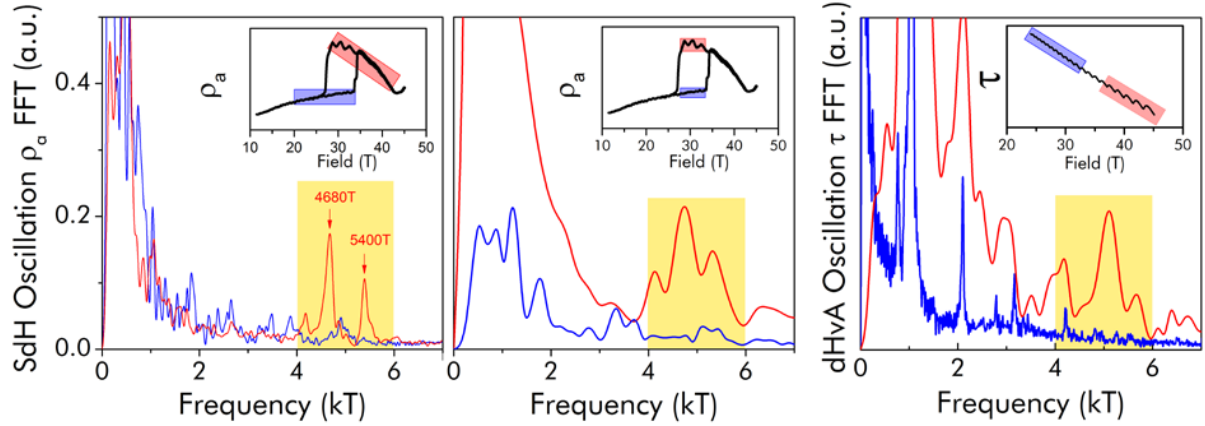

Comparison between the Shubnikov-de Haas frequencies observed in the in-plane transport across the hysteretic transition (left) as well as de Haas – van Alphen oscillations measured by magnetic torque (right) for 19° off H||c. Upon entering the high-resistance state (red), two new high frequencies appear and thus indicate a Fermi Surface reconstruction at the transition. The large hysteresis allows to directly compare the frequency spectrum of the upper branch to the lower branch within the same field window (middle). The high frequency oscillations appear over the whole hysteretic region on the upper branch and are completely absent in the lower branch. The appearance of the frequency can thus be clearly linked to the transition, and does not occur simply due to the enhancement of quantum oscillations in higher fields. The magnetic torque is a thermodynamic quantity and can thus shed light on the thermodynamic character of the transition. We observe the appearance of high frequency components also in the torque, in agreement with reference [20], indicating the thermodynamic nature of the transition despite the absence of a clear anomaly.

**Acknowledgements**: FIB and SEM work was supported by EMEZ and ScopeM at ETH Zurich. L.B. is supported by DOE-BES through award DE-SC0002613. We thank Bertram Batlogg, Brad Ramshaw, Joe Thompson, Suchitra Sebastian, Christoph Geibel and Michael Nicklas for interesting discussions and Philippe Gasser for supporting the FIB work. Work at Los Alamos was supported by the U.S. Department of Energy, Office of Science, Basic Energy Sciences, Materials Sciences and Engineering Division. The NHMFL facility is funded through the US NSF Cooperative Grant No. DMR-1157490, the DOE, and the State of Florida.

# Supplementary Material:

*S1– Device preparation*

All presented devices were extracted from macroscopic single crystals that were characterized by X-Ray and checked for signs of Indium impurities in a SQUID. The single crystals were then analyzed in a scanning electron microscope (SEM) for surface defects. A clean, scratch-free part of the crystal was selected and the crystal oriented using X-ray diffraction. In an initial step, a coarse (a,c) lamella with dimensions of ~ 60x60x(7-8) µm$^3$ was carved out of the crystal bulk with a 2.5nA, 30kV Ga$^{2+}$ focused ion beam (FEI Helios Nanolab 600i). The (a,b) face of the crystal was perpendicular to the ion beam during this process. In the next step, the sample was rotated in situ by 52°, to undercut the slice at 2.5nA, 30kV. In the last preparation step, the lamella was polished carefully using a lower current of 430pA, 30kV to remove potential damage layers from the high flux beam and to thin the lamella to its final thickness intended for the respective device (2-5 µm).

The lamella was then transferred onto a Si chip with photolithographically prepared Au leads, connecting the final structure to bonding pads. The crystal slice was extracted ex situ, and glued into the center of this chip using a small amount of epoxy glue. In a final procedure, the sample was reintroduced into the FIB, for further carving and electric contacts. The sample was contacted with ia-CVD deposited Pt in the FIB, at high ion flux to minimize resistance of the leads (2.5-9nA, 30kV). The typical resistance of each lead prepared this way was 20 Ohms. In addition, the square platelet was further cut into the shape shown in the main Figure 1 using low currents (430pA, 30kV). Preparing such a sample requires 4-5 full days of FIB work.

This microstructing procedure is essential for this experiment for four main reasons: 1) The low resistivity of CeRhIn$_5$ leads to very small signals in typically mm-sized as-grown crystals at the low currents required to avoid self-heating at low temperatures. After microstructuring the crystal into a long (~30µm) and thin (~5.5µm$^2$) bar, the resistance is greatly enhanced by the geometrical factor, facilitating high-precision measurements in the challenging environment of strong magnets. In addition, the FIB deposited contacts were previously shown to perform reliably in pulsed field conditions. 2) While the zero-field resistivity at low temperatures is only moderately anisotropic, significant and field-dependent anisotropy develops in magnetic fields. Therefore it is essential to ensure a homogeneous current profile in the sample to avoid field-dependent current redistribution due to changing anisotropy. The FIB microbars featuring side-contacts on a long bar are ideally suited to reliably distinguish between in- and out-of-plane resistivity. 3) Very high current densities can be achieved in such devices by applying only moderate currents due to the micron sized cross-section. While in principle low resistance contacts could sustain similar current densities with some self-heating, currents of a few Ampere will not be feasible in high magnetic fields due to the extreme Lorentz force acting on the wires. As CeRhIn5 is a good conductor, sustaining the depinning electric field $E_c$ requires a substantial current density that would be impractical in macroscopic crystals. 4) In micron sized samples, single domains may be isolated, allowing the study of intrinsic properties without the complications arising from domain dynamics.

**Figure S1 – Resistivity of the microstructured CeRhIn$_5$ devices**

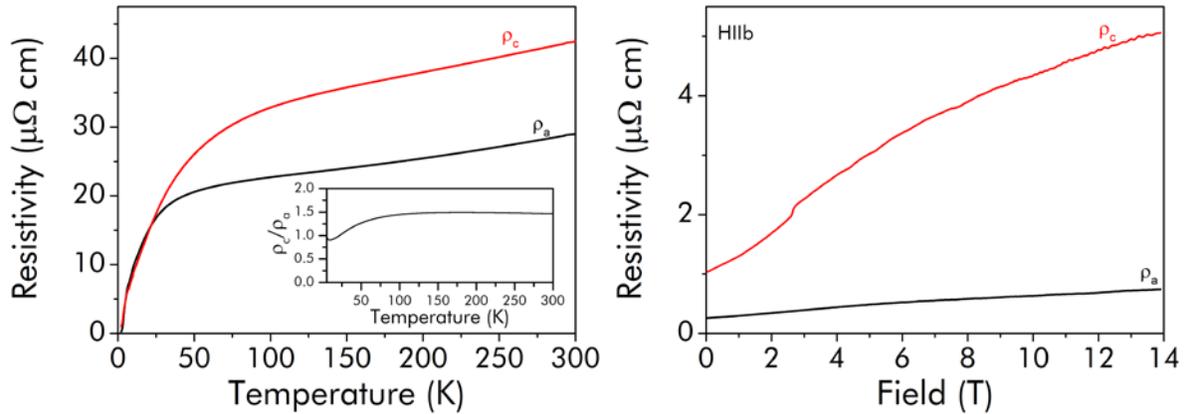

a) Temperature dependence of the resistivity along both crystal directions in zero field. The temperature dependence matches quantitatively to previous measurements in bulk CeRhIn$_5$ crystals. The high residual resistivity ratio and the low resistivities achieved at low temperatures indicate the high quality of the studied devices. Inset: Temperature dependence of the resistivity anisotropy.

b) Magnetoresistance along both crystallographic directions at T=1.9K and in fields up to 14T. The Shubnikov-de Haas oscillations are sensitive to lowest levels of impurities, and their observation and agreement with bulk samples demonstrates the high crystal quality of the device.

*S2 – Reproducibility*

The FIB technique used to microstructure the single crystals of CeRhIn$_5$ studied here involves high-energy ion irradiation and thus one has to take great care to investigate significant defects or changes to the material that may have been introduced by this preparation. We have found no sign of material degradation in any of the samples and in fact our current experimental evidence points towards a consistently high quality of such devices prepared out of heavy fermion compounds. We observed consistently high residual resistivity ratios in all prepared devices well above 200. The sample mainly used throughout this study showed residual ratios of ρ$_a$(300K)/ρ$_a$(80mK)~258 and ρ$_c$(300K)/ρ$_c$(80mK)~268. Another evidence for the high quality of our devices is the observation of Shubnikov-de Haas oscillations in fields as low as 4T. This oscillatory phenomenon originates from the quantization of electronic orbits on the Fermi surface into Landau levels and is known to be exceptionally sensitive to defects ($\omega_c\tau$~1-criterion)[22]. Its observation is generally accepted as strong evidence for very clean samples with a long mean-free-path. This leads us to conclude that the FIB process did not introduce significant scattering. Nine FIB-prepared samples of different shapes and sizes were produced and characterized in fields up to 14T, and all of them have shown Shubnikov-de Haas oscillations. Four of these were investigated in high dc- as well as in pulsed magnetic fields and all of them displayed the field-induced transition in quantitative agreement among them (See Figure S2 below).

Particularly the possibility of sample artifacts mimicking the reported density wave transition was carefully checked. During the FIB process some peculiar type of defect such as a micro-crack could have been introduced into the sample, causing extrinsic physical parameters such as field and temperature to conspire in a way to artificially create the observed resistivity jump. This scenario is highly unlikely, as the jump is accompanied by changes of the Shubnikov-de Haas spectrum, which indicates the thermodynamic nature of the transition. To nonetheless exclude such extrinsic phenomena, a total of 9 samples were prepared and 4 of them studied in high fields, in dc-fields at the National High Magnetic Field Laboratory in Tallahassee (NHMFL) as well as in pulsed fields in the pulsed field facility at Los Alamos National Laboratory (NHMFL-PFF). All studied samples consistently showed this transition at the same fields, however no hysteresis was observed in pulsed fields, indicating the highly metastable character of the lower branch. At high ramprates and high currents, the hysteresis also disappeared in dc-fields.

The Figure S2 shows the various sample designs used for this reproducibility check. In particular, care was taken to vary the sample geometry and mounting procedure to check for systematic sources of error. Highly different sample designs, from the U-shaped sample shown in the main manuscript, to thick and wide slabs, to thin, spring-like meander designs were fabricated. Two of them, CeRhIn5_#7 and CeRhIn5_H3 were mounted on epoxy glue, while the other two were not. These different types of mounting were prepared to check for the potential influence of strain originating from the differential thermal contraction. The consistent observation of Shubnikov – de Haas oscillations in the samples rules out significant strain, as inhomogeneous pressure smears out the frequencies and prevents their observation.

## Figure S2 – Reproducibility

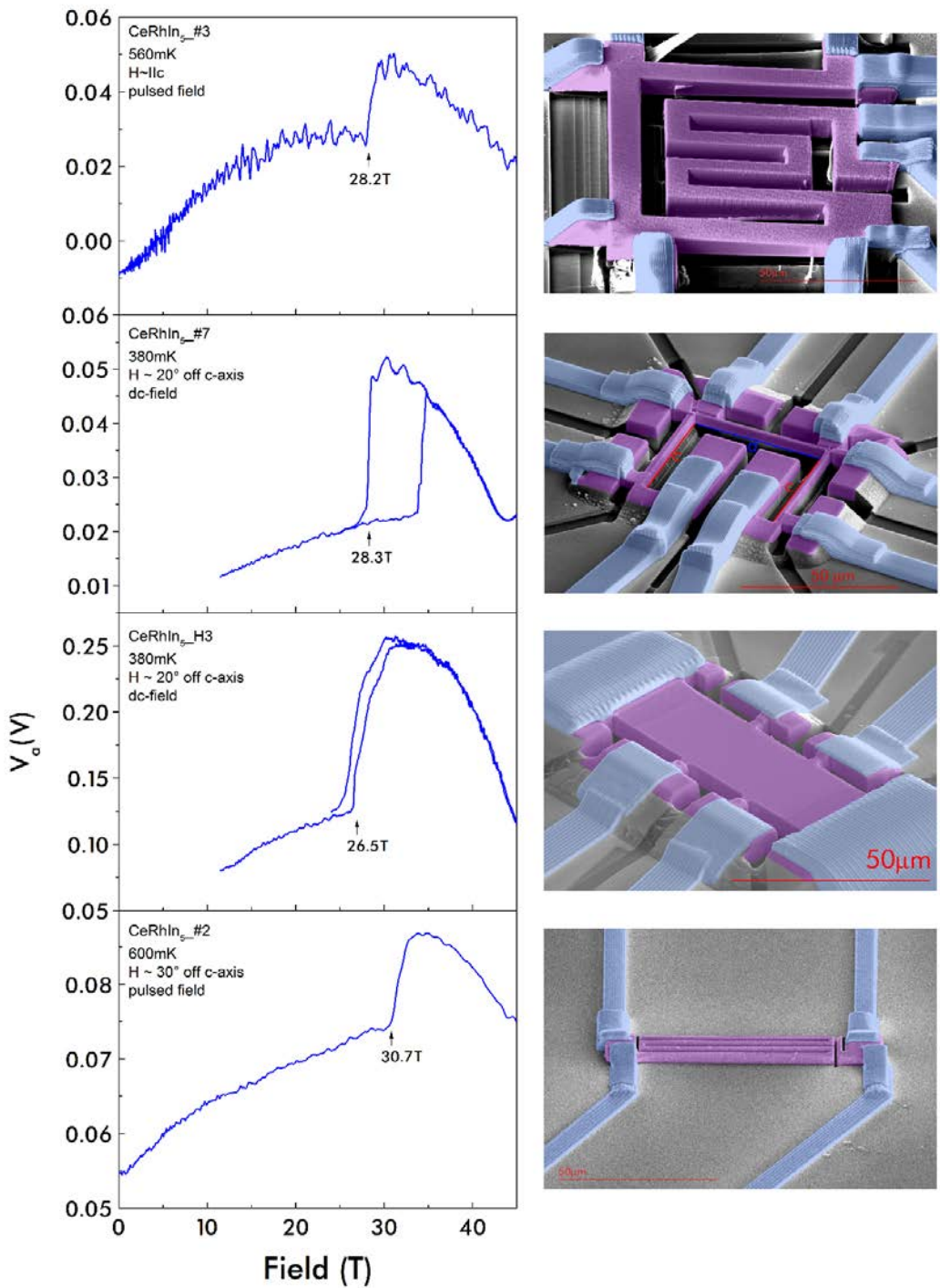

*S3 - Sample size influences hysteresis*

**Figure S3 – Comparison of a small and a large sample**

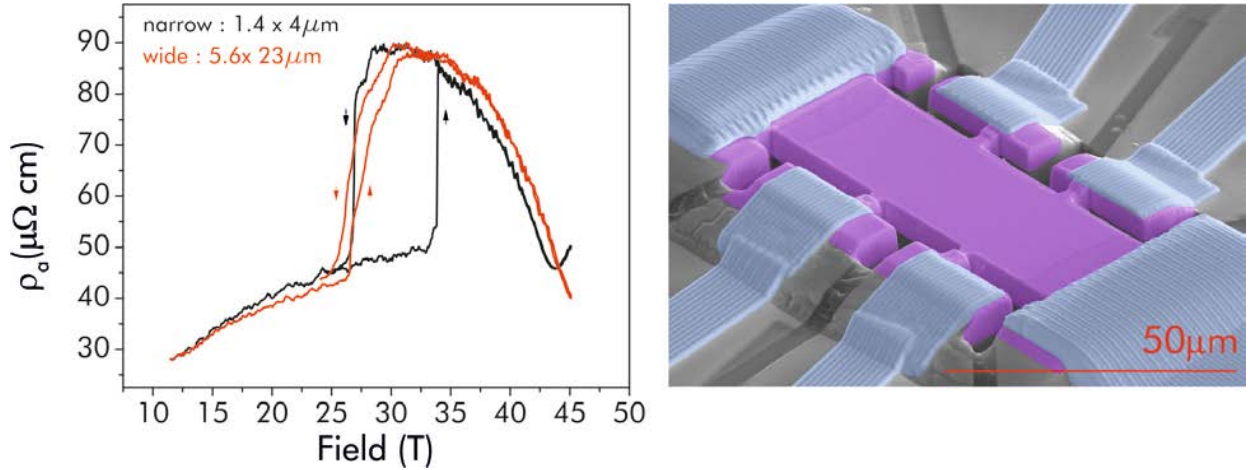

In-plane resistivity $\rho_a$ for the small sample whose data is displayed throughout this study (see Figure1, black trace) as well as $\rho_a$ from a larger device (red trace) shown on the right panel, both for fields aligned with the c-axis. The overall magnetoresistance behavior agrees well among both samples, however the hysteresis loop is significantly larger in the small sample. The charge-order transition for the large sample occurs at basically the same critical field as the one observed for the small sample when the field is swept down, thus indicating that this field of 27T is closer to the thermodynamically relevant field

While density-wave transitions are typically hysteretic, the observed hysteresis is surprisingly large. This opens up the exciting possibility to study the electronic matter with and without broken translational symmetry over a large range of magnetic field. A likely explanation for the metastability of the supercooled low-resistance branch is the micron-size of the sample. In most of our measurements on the small sample shown in the main Figure 1, the transitions between the low- and high-resistance states occur as step-like features, with the notable exception of a few upward jumps in which the system went into an intermediate resistance state (see for example Fig. S3, 33°). Charge-density-waves couple the electronic system with the crystal lattice, and are thus always accompanied by a small lattice distortion. The micron-sized cross-section and hence large surface-to-volume ratio may lead to a stronger mechanical coupling between the substrate and the microstructure compared to macroscopic samples, which prevents the release of strain at the charge order instability. Effectively, this would lead to an enhanced supercooling of the metallic state into the charge ordered phase. Such supercooled states are metastable, and slight disturbances such as fast field ramp rates or high currents could suddenly initiate the phase transition into the thermodynamically stable charge ordered phase. To check this hypothesis, a different sample of significantly larger dimensions (Figure S3, width: 24µm, thickness: 5.6 µm) was studied under the same conditions. In this larger sample, the transition occurs at a field value close to the drop in the smaller sample and it also shows hysteresis, however with smooth transitions in field instead of the step-like transitions (Figure S3). This suggests the lower-field transition to be closer to the true thermodynamic transition.

*S4 - Angular dependence of the critical field*

To further investigate the role of dimensionality, we have studied the angular dependence of the density wave transition by tilting the magnetic field in the (a,c) plane (Figure S4). The transition moves to higher field values as the field is rotated towards the a-direction. The upturn in angle does not follow a simple functional dependence; in particular it does not behave as $1/\cos(\theta)$ as one would expect if the out-of-plane field component would be the relevant parameter. Charge-density-waves in three-dimensional metals arise due to the nesting-enhancement of the susceptibility leading to a Fermi surface instability. One plausible scenario for a field-induced density wave transition would be a progressive enhancement of Fermi surface nesting as the field deforms it in the presence of anisotropic and band-dependent g-factors. Such a mechanism would result in a highly non-trivial angular dependence for the critical magnetic field with respect to the crystallographic axes. Above angles of 60°, the transition becomes unobservable. This is accompanied by a gradual reduction of the relative amplitude of the resistive jump, i.e. from its maximum at 20° to zero around 60°, indicating that at higher angles away from the c-axis the charge-order becomes energetically unfavorable.

**Figure S4 – Angular dependence**

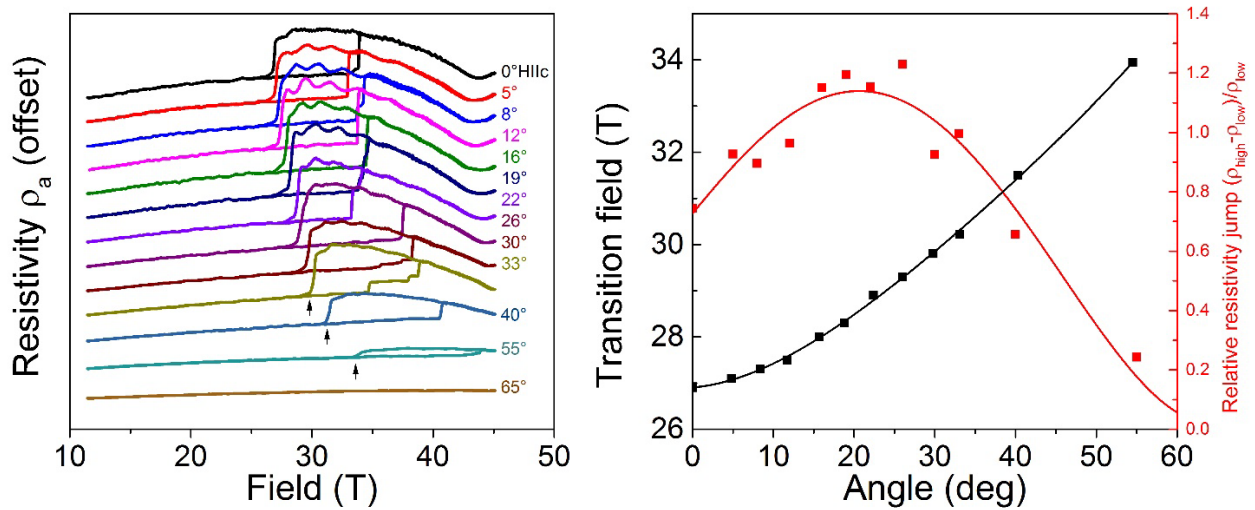

(left) In-plane resistivity as a function of the magnetic field at T=380mK, for various field angles relative to the c-axis, i.e. from 0° (=H∥c) to 65° as the field is rotated towards the b-direction. The field remained perpendicular to the applied current in a transverse magnetoresistance configuration at all studied angles. The drop at lower fields is always reproducible under all experimental conditions. The position in field of the jump from the low-resistance into the high resistance state at higher fields, however, depends on external conditions such as the applied current, field sweep rate and angle. The largest hysteresis was obtained by using slow field sweep rates (1T/min) and low currents (400$\mu$A). Nevertheless, jumps to intermediate resistivity values were observed, especially at higher angles.

(right) Field position of the drop from the high- into the low-resistance state (marked by arrows in a) as well as the height of the drop relative to the low-resistivity branch. The lowest transition field is observed for fields along the c-axis, and the transition field increases upon tilting the field towards the b-direction until $\theta \sim 60°$, above which no transition is observable. The relative height of the drop is difficult to define with precision due to the pronounced resistance oscillations in the high field state arising from the Shubnikov-de Haas effect. The relative drop is maximal at around 20° away from c-axis and decays upon further rotation of the field away from the c-axis, until the transition becomes undetectable at ~60°.